\pdfoutput=1
\documentclass[sigconf,compact]{acmart}
\usepackage{
    graphicx,
	url,
    hyperref,
    cleveref,
    float,
    multirow
}
\usepackage[acronym,nopostdot,nogroupskip,nonumberlist]{glossaries}
\glsdisablehyper
\def\BibTeX{{\rm B\kern-.05em{\sc i\kern-.025em b}\kern-.08emT\kern-.1667em\lower.7ex\hbox{E}\kern-.125emX}}
    
\copyrightyear{2022}
\acmYear{2022}
\setcopyright{ccby}
\acmConference[OSCAR 2022]{First Workshop on Open-Source Computer Architecture Research}{June 18, 2022}{New York City}
\acmDOI{}
\acmISBN{}

\begin{document}

\title{Muntjac -- Open Source Multicore RV64 Linux-capable SoC}

\author{Xuan Guo}
\email{Gary.Guo@cl.cam.ac.uk}
\affiliation[obeypunctuation=true]{%
  \institution{University of Cambridge},
  \country{UK}
}

\author{Daniel Bates}
\email{Daniel.Bates@cl.cam.ac.uk}
\affiliation[obeypunctuation=true]{%
  \institution{University of Cambridge},
   \country{UK}
}

\author{Robert Mullins}
\email{Robert.Mullins@cl.cam.ac.uk}
\affiliation[obeypunctuation=true]{%
  \institution{University of Cambridge},
   \country{UK}
}

\author{Alex Bradbury}
\email{asb@asbradbury.org}
\affiliation[obeypunctuation=true]{%
  \institution{lowRISC CIC},
  \country{UK}
}

\renewcommand{\shortauthors}{Guo, et al.}

\newacronym{RAS}{RAS}{return address stack}
\newacronym{BTB}{BTB}{branch target buffer}
\newacronym{ALU}{ALU}{arithmetic logic unit}
\newacronym{OS}{OS}{operating system}
\newacronym{PIPT}{PIPT}{physically-indexed physically-tagged}
\newacronym{CLINT}{CLINT}{core local interrupt}
\newacronym{PLIC}{PLIC}{platform-level interrupt controller}
\newacronym{LRSC}{LRSC}{load-reserved store-conditional}
\newacronym{LLC}{LLC}{last-level cache}

\newcommand{\muntjac}{Muntjac}

\graphicspath{{figures/}}

%

\maketitle

\section{Introduction}

Muntjac is an open-source\footnote{Available at https://github.com/lowRISC/muntjac under Apache-2.0 licence.} collection of components which can be used to build a multicore, Linux-capable system-on-chip. This includes a 64-bit RISC-V core, a cache subsystem, and TileLink \cite{tilelink} interconnect allowing cache-coherent multicore configurations. Each component is easy to understand, verify, and extend, with most being configurable enough to be useful across a wide range of applications. 

\begin{figure}[htbp!]
    \includegraphics[scale=0.6]{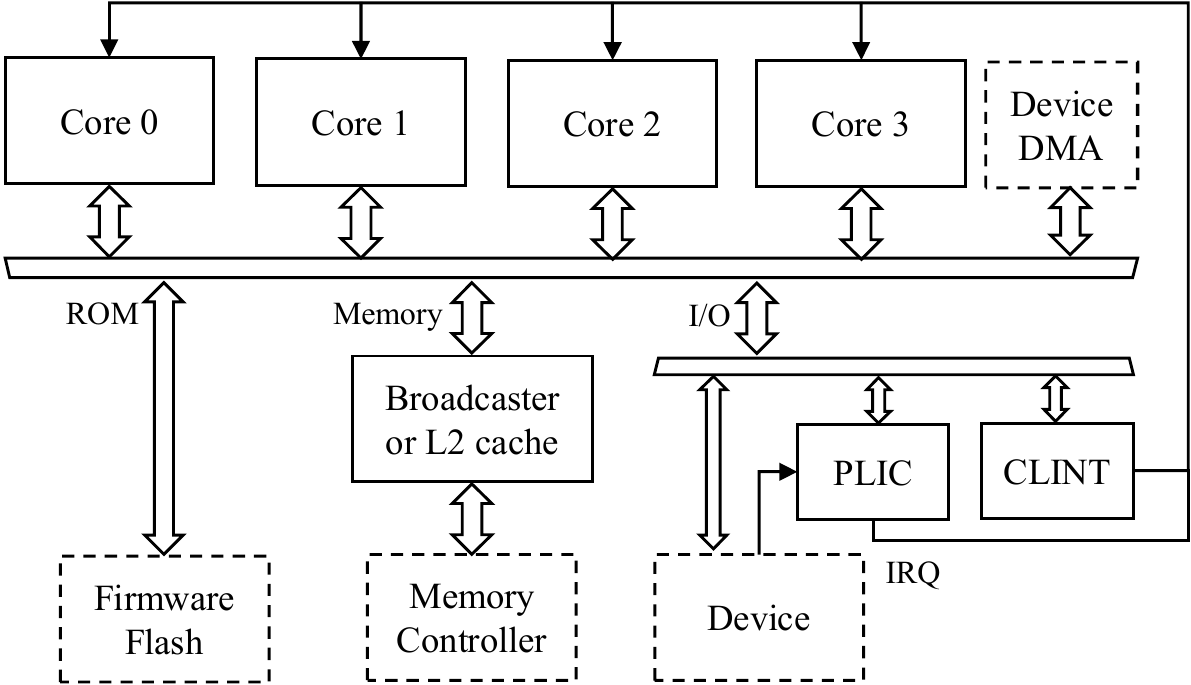}
    \caption{Example multi-core \muntjac{} system. All solid-border components are provided by \muntjac{}.}
    \label{fig:soc}
\end{figure}

Inspired by the lowRISC's Ibex project \cite{ibex}, we prioritise verification and documentation so users can make rapid progress and be confident that Muntjac's behaviour adheres to all appropriate standards. We focus on having clean, well-tested designs with clear routes to further customisation and improvements. Muntjac's performance is competitive with other open-source projects, and we aim for a power/performance/area compromise that maximises the value of Muntjac as a baseline design for others to use.


We anticipate Muntjac being valuable in the following situations, among others:
\begin{itemize}
  \item Education: the complete code for a high-quality modular processor is available, with documentation, verification support and examples, and lends itself to self-contained extensions. 
  \item Research: Muntjac can serve as a solid baseline for experimentation, removing the need to build and test huge swathes of general-purpose infrastructure needed to support a single novel component.
  \item Industry: many real-world applications value ``time to solution''. Muntjac is a reliable, configurable, complete starting point, allowing engineers to focus on the unique selling points of their target system.
\end{itemize}

Why use Muntjac? At the project's inception, there were a number of similar RISC-V projects available, but none had the right mix of features for us. We believe we have now reached a unique point in terms of performance, documentation, verification and ease of understanding. We have a consistent coding style, all components have been designed together to complement each other, and we are able to explain the design better to new contributors. Accessibility is aided by a relatively small and well-structured codebase, and usage of popular, well-supported hardware description languages (SystemVerilog) and standards (RISC-V, TileLink) to maximise interoperability with other tools and IP.
Everything ``just works'' out-of-the-box, and can be configured to suit new designs.

We aim to grow the community of Muntjac users, and together we will extend and improve the range of components we offer.
\section{Core Overview}

The Muntjac core is single-issue, in-order, scalar, and supports the RV64GC instruction set with machine and supervisor ISA. Floating-point extensions (F and D) are optional and can be configured with SystemVerilog parameters.

The Muntjac core is designed with modularity in mind. It aims to be easy to understand, verify and extend.
We aim to make components as loosely coupled as possible: in general, we choose distributed stall signals over a global stall signal or a stall-free design. Valid-ready signals are widely used both within components and between components, so each component and each pipeline stage can be mostly self-contained. Skid buffers are used when distributed stall signals start to cause timing issues.

\section{Pipeline Overview}

\begin{figure}[htbp!]
    \includegraphics[scale=0.5]{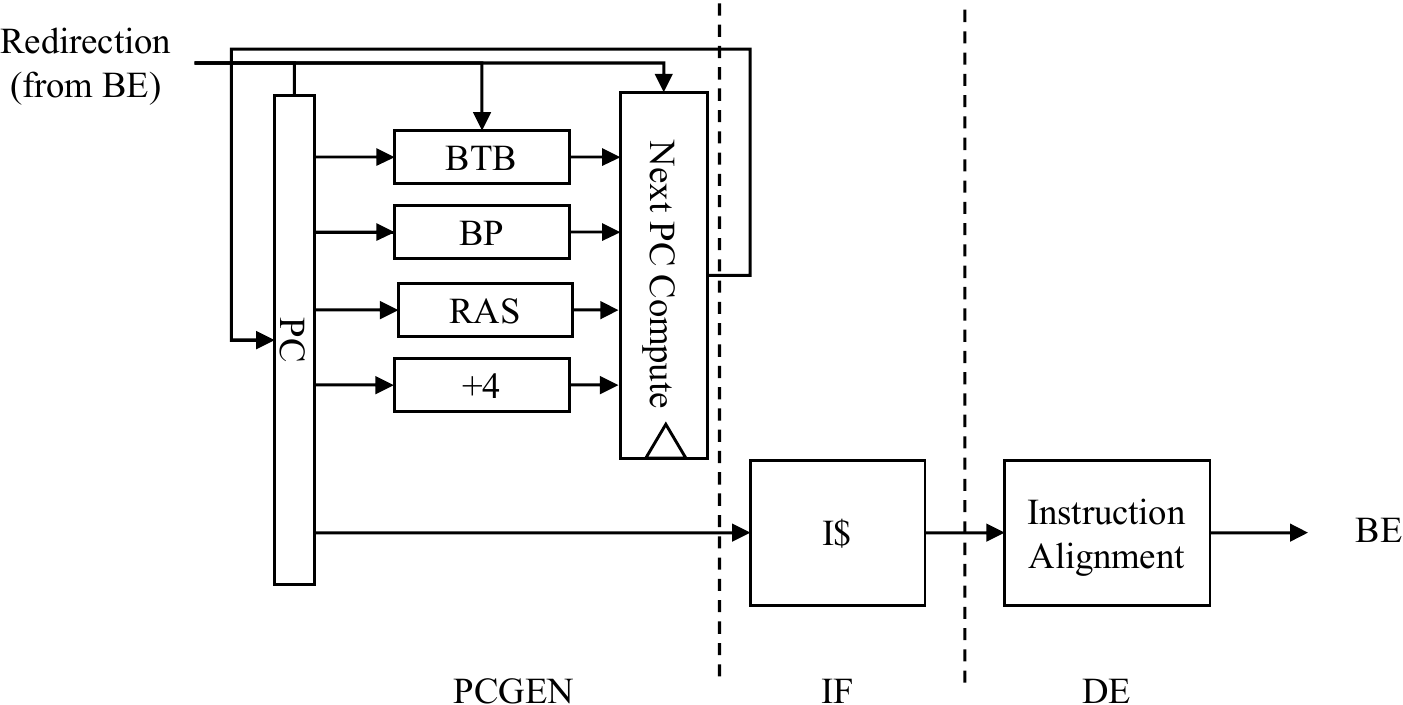}
    \caption{Muntjac frontend design}
    \label{fig:frontend}
\end{figure}

Muntjac's frontend design is illustrated in \Cref{fig:frontend}. There is a separate PCGEN stage that generates the next PC, without peeking into the fetched instruction bytes.

Muntjac supports the compressed instruction (C) extension for code size reduction. We do not provide a parameter to turn if off; this eliminates the misaligned instruction exception, which is the only exception that can possibly be generated by jump instructions, simplifying the backend.
The complexity of compressed instructions and misaligned instructions is entirely handled within the frontend; the instruction cache only needs to support accesses aligned to word (32-bit) boundaries.

The \gls{BTB}, \gls{RAS} and branch predictor are used to predict the next PC. Currently, the branch predictor is a simple bi-modal 2-bit saturating counter, but like other components, the interface between frontend and branch predictor is well-defined and a different implementation can be easily swapped in.

\begin{figure}[htbp!]
    \includegraphics[scale=0.45]{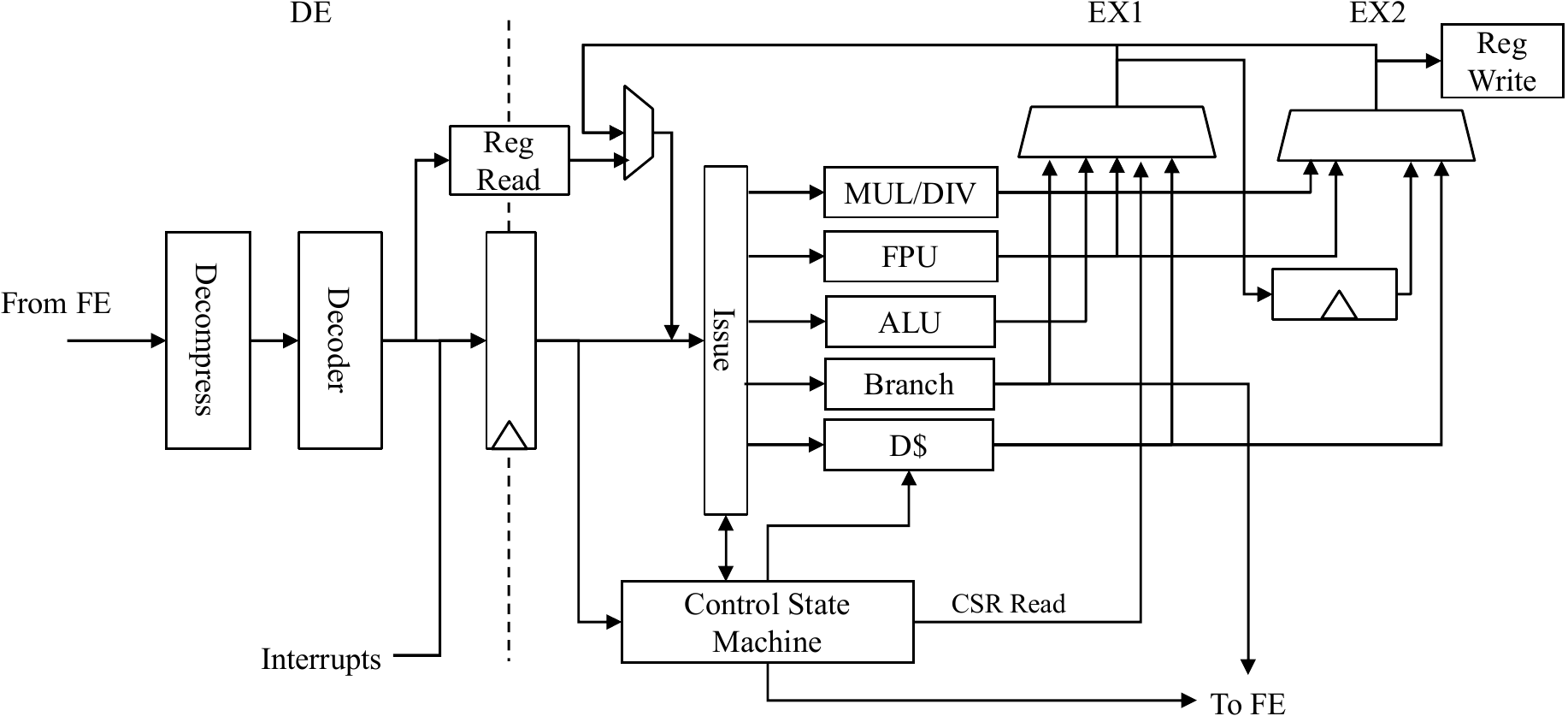}
    \caption{Muntjac backend design}
    \label{fig:backend}
\end{figure}

The backend is depicted in \Cref{fig:backend}. The issue logic manages control, data and structural hazards, and issues instructions into one of the functional units.

The \gls{ALU} and the branching unit complete in a single cycle, while the latency of other functional units may be multiple and variable number of cycles. To hide the latency of data cache accesses (usually 2 cycles when cache hits), there are two execution stages, EX1 and EX2. All instructions progress from EX1 to EX2 before their result is written back.

CSR accesses and other system instruction executions are handled outside the normal data flow; as they can affect global states they can only execute while the pipeline is empty.

An optional FPU is available with \muntjac{}. Unlike many other implementations, we opted not to use a recoded format internally, so that NaN-boxing behaviour is more easily implemented and hard-to-test bugs related to recoding can be avoided.
We provide an option that enables FP registers (and FP load/store instructions) but disables the FPU. This mode allows efficient emulation of a RV64GC core without bearing the cost of the FPU: M-mode firmware can trap and emulate the floating-point operations, but trapping is not needed when the \gls{OS} merely saves or restores FP contexts without performing actual computations.
\section{Cache Overview}

\muntjac{} provides cache implementations with interfaces that comply with the TileLink protocol. Both instruction and data caches communicate with the pipeline via generic valid-ready interfaces, so they have to manage the request replays themselves without help from the frontend/backend.

\muntjac{}'s data cache implements the TileLink Cached (TL-C) protocol for its memory-facing interface and is therefore multi-core-capable. The cache is set-associative, and for each way tags and data are stored in separate single-port SRAMs. Tag SRAM, data SRAM and TLB lookup all take place in parallel when an access request arrives from the backend. The cache has an AMOALU independent from backend's ALU for atomic operations. Forward progress is guaranteed by locking cache lines for 16 cycles after a refill.

\muntjac{}'s instruction cache is similar to the data cache but with narrow access and write support removed, and width reduced to 32-bit. It implements the TileLink Uncached Heavy (TL-UH) protocol.

\muntjac{} provides a range of TileLink IPs allowing individual \muntjac{} cores to be aggregated to a multi-core system. For simple designs, a broadcaster can be used to ensure data L1s are coherent. A reference directory-based L2 cache design is also provided. This cache uses TL-C for both the CPU-facing link and memory-facing link, so despite its name, it could be cascaded and used as L3 caches as well. This cache design can also be used as individual banks of a larger cache. Various adapter and bridge IPs are provided to convert TL-C to TL-UH and from TL-UH to AXI.

\section{Utilisation and Performance}

For utilisation and performance numbers, we use the default configuration parameters. The L1 data and instruction cache are 16 KiB each, 4-way associative. The L1 data and instruction TLBs are 32 entry each and 4-way associative. The L2 cache is 64KiB in size and also 4-way associative.

When targeting Xilinx Kintex 7 with Synopsys Synplify, with FPU on, the core uses 17420 LUTs and 6683 registers (13663 LUTs and 4538 registers for the pipeline) at 89MHz; with FPU off, the core uses 9902 LUTs and 5266 registers (6022 LUTs and 3098 registers for the pipeline alone) at 97MHz; with FP registers but no FPUs, the core uses 9924 LUTs and 5281 registers (6111 LUTs and 3121 registers for the pipeline) at 95MHz.
\muntjac{} attempts to be both FPGA and ASIC friendly and we did not employ FPGA-specific optimisations.

On GCC 9.2.0, with optimisation and tuning turned on, \muntjac{} achieves Dhrystone score of 2.17 DMIPS/MHz and CoreMark score of 3.01/MHz.

\section{Verification}

We have put considerable effort into verification given its importance for encouraging adoption of the IP and growing its user base; we believe it can empower users by simplifying the process of validating modifications and extensions.

We validate the Muntjac core using a combination of systematic tests (\texttt{riscv-tests} \cite{riscv-tests}) and random test generation (\texttt{riscv-dv} \cite{riscv-dv}). We also regularly run ad-hoc system tests such as booting Debian Linux and running parallel benchmarks such as PARSEC \cite{parsec}. We test our TileLink IP using an assertion suite, a set of functional coverpoints, and a custom random traffic generator.

\section{Acknowledgement}

This work was kindly supported by lowRISC CIC.

%
%

%

\bibliographystyle{ACM-Reference-Format}
\bibliography{references}

\end{document}